\documentclass[a4paper,11pt]{article}
\usepackage[hmargin=1.5cm,vmargin=3cm]{geometry,}
\usepackage{graphicx, amssymb, textcomp, hyperref, amsmath}
\title{\bf Exact solutions to relativistic singular fractional power potentials}
\author{\Large Davids Agboola \footnote{d.agboola@maths.uq.edu.au}~ and~ Yao-Zhong Zhang \footnote{yzz@maths.uq.edu.au}}
\date{\it School of Mathematics and Physics, The University of Queensland, \\Brisbane, QLD 4072, Australia}
\begin{document}
\maketitle
\vspace{0.5in}
\noindent {\bf Abstract:} We present (exact) solutions of the Dirac equation with equally mixed interactions for a single fermion bounded by the family of fractional power singular potentials. Closed-form expressions as well as numerical values for the energies were obtained. The wave functions and the allowed values of the potential parameters for the first two members of the family are obtained in terms of a set of algebraic equations. The non relativistic limit is also discussed and using the Hellmann-Feynmann theorem, some useful expectation values are obtained.   

\vspace{.3in}



\vskip.5in
\section{Introduction} 
The family of singular potentials appears to be very significant in many aspects of modern physics as an extensive literature has been developed on the subject (see for example \cite{Case1950}-\cite{DDA12} and the references therein). One of the early works that generated much interest in the study of singular potentials was presented in \cite{PR1962} 
where it was argued that real world interactions were likely to be highly singular and thus
the study of singular potentials rather than regular potentials might be more relevant physically. 
This was thereafter followed by applications of singular potentials to the study of gaseous ions moving through a gas 
\cite{VW1954} and also in the investigation of the elastic differential cross sections for high energy scattering 
\cite{OB81, GE98, PW1964, Dolinszky1980}, magnetic resonances between massless and massive spin-$\frac{1}{2}$ particles \cite{AOB80}, decay rate in mesons \cite{XS91}, interatomic or intermolecular diatomic molecues \cite{IK2007,ES1984} and non-polar molecules \cite{DV1987,SD2010}.

One particular class of the power-law potentials is the singular fractional power potentials, which are power-law potentials with rational powers. This class of potentials has recently found applications in models exhibiting shape resonance behaviour \cite{XS91} and in the description of quark-antiquark interations \cite{FHS79}. 
Bound state solutions of relativistic and non-relativistic models with such potentials cannot be analytically obtained, as the models are not exactly solvable. Thus, these models have  been discussed via various approximation and computational methods \cite{BS20,AS02,PB2012,BDK95}. The methods used in most previous works were ill motivated, without a well defined solution structure. As a result, closed form representation expression for the general solutions have not been previously obtained.

Recently, the functional Bethe ansatz method \cite{Zhang11} has been used to obtain the (exact) solutions to the non-relativistic quantum mechanical model with integer power singular potentials proposed in \cite{DZ12c,AZ11}. The aim of this paper is to solve the Dirac equation with the singular fractional power-law potentials of the form
\begin{equation}\label{eq:9}
V_N(r)=\sum_{p=1}^{2N-1}\frac{a_p}{r^{p/N}},~~~~~2\leq N\in \mathbb{N},
\end{equation} where $r\in (0,\infty)$ and $a_p$ are real parameters. Obviously, $V_1(r)$ is the well known Coulomb interaction, which will not be considered in our further discussion except as a limiting case.
In particular we obtain exact, closed form solutions of the Dirac equation for the $N=2$ and $N=3$ cases using the Bethe ansatz method. 

This work is  organized as follows. Section 2 deals with the transformation of the Dirac equation with the singular fractional power-law potentials into a QES second order differntial equation. Sections 3 and 4 present exact closed-form polynomial 
solutions to the square-root singular potential ($N=2$) and the third-order singular potential ($N=3$) respectively. 
With appropriate limits, the non relativistic spectrum and wavefunction for the  square-root singular potential ($N=2$) are obtained in section 5. In addition, using the Hellmann-Feynmann theorem, we worked out some expectation values are for the model.  Finally, in section 6 we provide some concluding remarks.
 
\section{Dirac equation with fractional singular potentials}
The Dirac equation for a single fermion with mass $\mu$ moving in a spherically symmetric central scalar $S(r)$ and vector $V(r)$ potentials can be written in natural units $\hbar=c=1$ as \cite{WG81}
\begin{equation}\label{eq:1}
H\Psi{({\bf r})}=E_{n_r\kappa}\Psi({\bf r})\hspace{.2in}\mbox{where}\hspace{.1in}H=\sum_{j=1}^3\hat{\alpha}_jp_j+\hat{\beta}[\mu+S(r)]+V(r) 
\end{equation} and $E_{n\kappa}$ is the relativistic energy, $\{\hat{\alpha}_j\}$ and $\hat{\beta}$ are Dirac matrices defined as 

\begin{equation}\label{eq:2}
\hat{\alpha}_j=\begin{pmatrix}
0&\hat{\sigma}_j \\
\hat{\sigma}_j&0

\end{pmatrix}\hspace{.3in}\hat{\beta}=\begin{pmatrix}
\bf{1}&0 \\
0&\bf{-1}

\end{pmatrix}
\end{equation}
where $\hat{\sigma}_j$ is the Pauli's $2\times 2$ matrices and $\mathbf{1}$ is a 2x2 unit matrix, which satisfy anti-commutation relations
\begin{equation}\label{eq:3}
\begin{array}{lrl}
\hat{\alpha}_j\hat{\alpha}_k+\hat{\alpha}_k\hat{\alpha}_j&=&2\delta_{jk}\bf{1}\\
\hat{\alpha}_j\hat{\beta}+\hat{\beta}\hat{\alpha}_j&=&0\\
{\hat{\alpha}_j}^2=\hat{\beta}^2&=&\bf{1}

\end{array}
\end{equation}
and $p_j$ is the three momentum which can be written as
\begin{equation}\label{eq:4}
p_j=-i\partial_j=-i\frac{\partial}{\partial x_j} \hspace{.2in} 1\leqslant j\leqslant 3.
\end{equation}
The orbital angular momentum operators $L_{jk}$, the spinor opertaors $S_{jk}$ and the total angular momentum operators $J_{jk}$ can be defined as follows:
\begin{equation}\label{eq:5}
L_{jk}=-L_{jk}=ix_j\frac{\partial}{\partial x_k}-ix_k\frac{\partial}{\partial x_j},\hspace{.2in} S_{jk}=-S_{kj}=i\hat{\alpha}_j\hat{\alpha}_k/2,\hspace{.2in} J_{jk}=L_{jk}+S_{jk},$$$$
L^2=\sum_{j<k}^3L^2_{jk},\hspace{.2in}S^2=\sum_{j<k}^3S^2_{jk},\hspace{.2in}J^2=\sum_{j<k}^3J^2_{jk}, \hspace{.2in} 1\leqslant j< k\leqslant 3.
\end{equation}
For a spherically symmetric potential, total angular momentum operator $J_{jk}$ and the spin-orbit operator $\hat{K}=-\hat{\beta}(J^2-L^2-S^2+1/2)$ commutate with the Dirac Hamiltonian. For a given total angular momentum $j$, the eigenvalues of $\hat{K}$ are $\kappa=\pm(j+1/2)$; $\kappa=-(j+1/2)$ for aligned spin $j=\ell+\frac{1}{2}$ and $\kappa=(j+1/2)$ for unaligned spin $j=\ell-\frac{1}{2}$. Moreover, the spin-orbital quantum number $\kappa$ is related to the orbital angular quantum number $\ell$ and the pseudo-orbital angular quantum number $\tilde{\ell}=\ell+1$ by the expressions $\kappa(\kappa+1)=\ell(\ell+1)$ and $\kappa(\kappa-1)=\tilde{\ell}(\tilde{\ell}+1)$ respectively for $\kappa=\pm 1, \pm 2,\dots$. The spinor wave functions can be classified according to the radial quantum number $n_r$ and the spin-orbital quantum number $\kappa$ and can be written using the Dirac-Pauli representation
\begin{equation}\label{eq:6}
\Psi_{n_r\kappa}({\bf r})=\frac{1}{r}\left(\begin{array}{lll}
F_{n_r\kappa}(r)Y_{jm}^\ell\left(\theta,\phi\right)\\\\
iG_{n_r\kappa}(r)Y^{\tilde{\ell}}_{jm}\left(\theta,\phi\right)
\end{array}\right)
\end{equation}
where $F_{n_r\kappa}(r)$ and $G_{n_r\kappa}(r)$ are the radial wave function of the upper- and the lower-spinor components respectively, $Y_{jm}^\ell\left(\theta,\phi\right)$ and $Y^{\tilde{\ell}}_{jm}\left(\theta,\phi\right)$ are the spinor spherical harmonic functions coupled with the total angular momentum $j$. The orbital and the pseudo-orbital angular momentum quantum numbers for spin symmetry  $\ell$ and  and pseudospin symmetry $\tilde{\ell}$ refer to the upper- and lower-component respectively.  
Substituting  Eq.\,\eqref{eq:6} into  Eq.\,\eqref{eq:1}, and seperating the variables we obtain the following coupled radial Dirac equation for the spinor components:
\begin{equation}\label{eq:7}
\left(\frac{d}{dr}+\frac{\kappa}{r}\right)F_{n_r\kappa}(r)=[\mu+E_{n_r\kappa}-\Delta(r)]G_{n_r\kappa}(r),\\
\end{equation}
\begin{equation}\label{eq:8}
\left(\frac{d}{dr}-\frac{\kappa}{r}\right)G_{n_r\kappa}(r)=[\mu-E_{n_r\kappa}+\Sigma(r)]F_{n_r\kappa}(r).
\end{equation}
where $\Sigma(r)=V(r)+S(r)$, $\Delta(r)=V(r)-S(r)$ and $n_r$ is the radial quantum number. For an equally mixed interaction, $\Delta(r)=0$, thus if we differentiate  Eq.\,\eqref{eq:7}, followed by the insertion of Eqs.\,\eqref{eq:8} and \eqref{eq:9}, we arrive at following Schr\"odinger-like second order differential equation
\begin{equation}\label{eq:10a}
F''_{n_r\kappa}(r)+\left[\left(E^2_{n_r\kappa}-\mu^2\right)-V_{\mbox{\footnotesize eff}}(r)\right]F_{n_r\kappa}(r)=0,
\end{equation} where the effective potential $V_{\mbox{\footnotesize eff}}(r)$ is given as
\begin{equation}\label{eq:10b}
V_{\mbox{\footnotesize eff}}(r)=\frac{\kappa(\kappa+1)}{r^2}+2(\mu+E_{n_r\kappa})\sum_{p=1}^{2N-1}\frac{a_p}{r^{p/N}}=
\frac{\kappa(\kappa+1)}{r^2}+2(\mu+E_{n_r\kappa})\left[\sum_{s=1}^{N}\frac{(-1)^{2N+s}a_s}{r^{s/N}}+\sum_{q=N+1}^{2N-1}\frac{a_q}{r^{q/N}}\right]
.
\end{equation}
If all $a_p$ are positive, then the effective potential $V_{\mbox{\footnotesize eff}}(r)$ becomes monotonously decreasing, which leads to no bound state \cite{LR83,SZ84,RH1986}. However, by alternating the signs before the first $N$ terms, the potential becomes bounded (as shown in Fig 1) and thus we may determine the possible bound states. Moreover, for numerical computation, the parameters $a_s$ can be made free while $a_q$ must be constrained for the systems to possess exact solutions. 

With the transformation 
\begin{equation}\label{eq:11}
F_{n_r,\kappa}(r)=x^{N(\kappa+1)}e^{w(x)}y(x),~~~~~w(x)=\sum_{p=1}^Nb_px^p,~~~~x=r^{1/N},
\end{equation}
Eq.\,\eqref{eq:10a} becomes 
\begin{equation}\label{eq:12}
\small\left\{\frac{x}{N^2}\frac{d^2}{dx^2}+\left[\frac{1-N}{N^2}+\frac{2xw'}{N^2}+\frac{2(\kappa+1)}{N}\right]\frac{d}{dx}+\left[(E^2_{n_r\kappa}-\mu^2)x^{2N-1}-2(\mu+E_{n_r\kappa})\sum_{p=1}^{N}{a_p}{x^{2N-1-p}}\right.\right.$$$$\left.\left.
+\frac{(1-N)w'}{N^2}+\frac{(w'^2+w'')x}{N^2}+\frac{2(\kappa+1)w'}{N}\right]\right\}y(x)=0.
\end{equation}
By suitably choosing the parameters $b_p$, which appears in the exponential prefactor, one can obtain exact solutions to  Eq.\,\eqref{eq:12}, provided the potential parameters satisfy some certain conditions. This is an essential feature of a quasi-exactly solvable system which can also be attributed to the quantization procedure of the system \cite{LBC12}. In what follows, we examine the $N=2$ and $N=3$ cases using the Bethe ansatz method.   

\section{Square-root singular potential ($N=2$)}
In this case, the interaction becomes 
\begin{equation}\label{eq:13}
V_{2}(r)=-\frac{a_1}{\sqrt{r}}+\frac{a_2}{r}+\frac{a_3}{r^{\frac{3}{2}}},~~a_1,a_2>0
\end{equation}
and with an appropriate choice of parameter $b_p$, Eqs.\,\eqref{eq:11} and \eqref{eq:12} reduce to
\begin{equation}\label{eq:14}
F_{n_r\kappa}(r)=x^{2(\kappa+1)}~e^{ w(x)}y(x),~~~~w(x)=-x^2(\mu^2-E_{n_r\kappa}^2)^{\frac{1}{2}}+\frac{2a_1 x(\mu+E_{n_r\kappa})}{\left({\mu^2-E_{n_r\kappa}^2}\right)^{\frac{1}{2}}},~~~~x=r^{\frac{1}{2}},
\end{equation} and 
\begin{equation}\label{eq:15}
xy''(x)+\left[-4x^2(\mu^2-E_{n_r,\kappa}^2)^{\frac{1}{2}}+\frac{4xa_1(\mu+E_{n_r,\kappa})}{({\mu^2-E_{n_r,\kappa}^2})^{\frac{1}{2}}}+(4\kappa+3)\right]y'(x)+4x\left[\frac{a_1^2(\mu+E_{n_r\kappa})^2}{(\mu^2-E_{n_r\kappa}^2)}\right.$$$$
\left.-2(\kappa+1)(\mu^2-E_{n_r\kappa}^2)^\frac{1}{2}-2a_2(\mu+E_{n_r\kappa})\right]y(x)=8(\mu+E_{n_r\kappa})\left[a_3-\frac{\left(\kappa+\frac{3}{4}\right)a_1}{(\mu^2-E_{n_r\kappa}^2)^{\frac{1}{2}}}\right]y(x).
\end{equation} respectively. Although, the Klein-Gordon case for the above model has been discussed in \cite{PB2012} using the $sl_2$ algebraization, however, in what follows, we show that Eq.\,\eqref{eq:15} is quasi-exactly solvable and therefore possesses polynomial solutions of degree $n_r\geq 0$, which we write in the form
\begin{equation}\label{eq:16}
y(x)=\prod_{i=1}^{n_r}(x-x_i),~~~~~~~~y(x)\equiv 1~~\mbox{for}~~n_r=0, 
\end{equation}
where $\{x_i\}$ are the roots of the polynomials to be determined. To solve equation \eqref{eq:15}, we apply the functional Bethe ansatz method. Substituting \eqref{eq:16} into \eqref{eq:15}, we obtain
 \begin{equation}\label{eq:17}
\sum_{i=1}^{n_r}\frac{x}{x-x_i}\sum_{j\neq i}^{n_r}\frac{2}{x_i-x_j}+\left[-4x^2(\mu^2-E_{n_r\kappa}^2)^{\frac{1}{2}}+\frac{4xa_1(\mu+E_{n_r\kappa})}{({\mu^2-E_{n_r\kappa}^2})^{\frac{1}{2}}}+(4\kappa+3)\right]\sum_{i=1}^{n_r}\frac{1}{x-x_i}~~~~$$$$
~~~~~+4x\left[\frac{a_1^2(\mu+E_{n_r\kappa})^2}{(\mu^2-E_{n_r\kappa}^2)}-2(\kappa+1)(\mu^2-E_{n_r\kappa}^2)^\frac{1}{2}-2a_2(\mu+E_{n_r\kappa})\right]=8(\mu+E_{n_r\kappa})\left[a_3-\frac{\left(\kappa+\frac{3}{4}\right)a_1}{(\mu^2-E_{n_r\kappa}^2)^{\frac{1}{2}}}\right]
\end{equation}
The right hand side of this equation is a constant, while the left hand side is a meromorphic 
function with simple poles $x=x_i$ and singularity at $x=\infty$. For this equation to be valid, 
the left hand side must also be a constant. We thus demand that the coefficients of the powers of 
$x$ as well as the residues at the 
simple poles of the left hand side be zero. Following Liouville's theorem, this is the necessary and sufficient condition  for the left hand side of \eqref{eq:17} to be a constant.

 Executing this demand, we have the following transcendental energy equation
\begin{equation}\label{eq:18}
\left[n_r+2(\kappa+1)\right]\left(\mu^2-E_{n_r\kappa}^2\right)^\frac{1}{2}-\frac{a_1^2(\mu+E_{n_r\kappa})^2}{(\mu^2-E_{n_r\kappa}^2)}+2a_2(\mu+E_{n_r\kappa})=0,
\end{equation} and the corresponding spinor wavefunction
\begin{equation}\label{eq:18b}\footnotesize
F_{n_r\kappa}(r)\sim x^{2(\kappa+1)}\left[\prod_{i=1}^{n_r}(x-x_i)\right]\exp\left[-x^2(\mu^2-E_{n_r\kappa}^2)^{\frac{1}{2}}+\frac{2a_1 x(\mu+E_{n_r\kappa})}{\left({\mu^2-E_{n_r\kappa}^2}\right)^{\frac{1}{2}}}\right], ~~~~x=r^{\frac{1}{2}}$$$$
\footnotesize G_{n_r\kappa}(r)\sim x^{2\kappa}\left[\prod_{i=1}^{n_r}(x-x_i)\right]\exp\left[-x^2(\mu^2-E_{n_r\kappa}^2)^{\frac{1}{2}}+\frac{2a_1 x(\mu+E_{n_r\kappa})}{\left({\mu^2-E_{n_r\kappa}^2}\right)^{\frac{1}{2}}}\right]\times~~~~~~~~~~~~~~~~ $$$$~~~~~~~~~~~~~~~~~~~~~~~~~\frac{1}{\mu+E_{n_r\kappa}}\left\{-x^2(\mu^2-E_{n_r\kappa}^2)^{\frac{1}{2}}+x\left(\frac{a_1 (\mu+E_{n_r\kappa})}{\left({\mu^2-E_{n_r\kappa}^2}\right)^{\frac{1}{2}}}+\frac{1}{2}\sum_{i=1}^{n_r}\frac{1}{x-x_i}\right)+2\kappa+1\right\}, 
\end{equation}
subject to the constraint
\begin{equation}\label{eq:19}
8(\mu+E_{n_r\kappa})\left[a_3-\frac{\left(\kappa+\frac{3}{4}\right)a_1}{(\mu^2-E_{n_r\kappa}^2)^{\frac{1}{2}}}\right]=-4\left(\mu^2-E_{n_r\kappa}^2\right)^\frac{1}{2}\sum^{n_r}_{i=1}x_i+n_r(4\kappa+3),
\end{equation}
where the roots $\{x_i\}$ satisfy the Bethe ansatz equations
\begin{equation}\label{eq:20}
\sum_{j\neq i}^{n_r}\frac{1}{x_i-x_j}=2x_i(\mu^2-E_{n_r\kappa}^2)^{\frac{1}{2}}-\frac{2a_1(\mu+E_{n_r\kappa})}{({\mu^2-E_{n_r\kappa}^2})^{\frac{1}{2}}}-\frac{(4\kappa+3)}{2x_i},\hspace{0.2in}i=1,\dots, n_r.
\end{equation}
It is easy to see that $y(x)=1$ is a solution of \eqref{eq:15} for certain values of the potential parameters.
This solution corresponds to the $n_r=0$ case in the general expressions above.  Thus from  Eqs.\,\eqref{eq:18} and \eqref{eq:18b} we have the ground state solutions
\begin{equation}\label{eq:21}
2(\kappa+1)\left(\mu^2-E_{0,\kappa}^2\right)^\frac{1}{2}-\frac{a_1^2(\mu+E_{0,\kappa})^2}{(\mu^2-E_{0,\kappa}^2)}+2a_2(\mu+E_{0,\kappa})=0,
~~~~~~\kappa=1,2,\dots\end{equation} and wavefunctions
\begin{equation}\label{eq:22}\footnotesize
F_{0,\kappa}(r)\sim x^{2(\kappa+1)}\exp\left[-x^2(\mu^2-E_{0,\kappa}^2)^{\frac{1}{2}}+\frac{2a_1 x(\mu+E_{0,\kappa})}{\left({\mu^2-E_{0,\kappa}^2}\right)^{\frac{1}{2}}}\right], ~~~~x=r^{\frac{1}{2}}$$$$
\footnotesize G_{0,\kappa}(r)\sim x^{2\kappa}\exp\left[-x^2(\mu^2-E_{0,\kappa}^2)^{\frac{1}{2}}+\frac{2a_1 x(\mu+E_{0,\kappa})}{\left({\mu^2-E_{0,\kappa}^2}\right)^{\frac{1}{2}}}\right]\times~~~~~~~~~~~~~~~~ $$$$~~~~~~~~~~~~~~~~~~~~~~~~~\frac{1}{\mu+E_{0,\kappa}}\left\{-x^2(\mu^2-E_{0,\kappa}^2)^{\frac{1}{2}}+\frac{a_1 x(\mu+E_{0,\kappa})}{\left({\mu^2-E_{0,\kappa}^2}\right)^{\frac{1}{2}}}+2\kappa+1\right\}, 
\end{equation}
where
\begin{equation}\label{eq:22b}
a_3=\frac{\left(\kappa+\frac{3}{4}\right)a_1}{(\mu^2-E_{0,\kappa}^2)^{\frac{1}{2}}}
.\end{equation}
Similarly, the first excited state solutions corresponding to $n_r=1$ are given as
\begin{equation}\label{eq:23}
(2\kappa+3)\left(\mu^2-E_{1,\kappa}^2\right)^\frac{1}{2}-\frac{a_1^2(\mu+E_{1,\kappa})^2}{(\mu^2-E_{1,\kappa}^2)}+2a_2(\mu+E_{1,\kappa})=0,~~~~~~\kappa=1,2,\dots
\end{equation} and the corresponding wavefunction
\begin{equation}\label{eq:24}\footnotesize
F_{1,\kappa}(r)\sim x^{2(\kappa+1)}(x-x_1)\exp\left[-x^2(\mu^2-E_{1,\kappa}^2)^{\frac{1}{2}}+\frac{2a_1 x(\mu+E_{1,\kappa})}{\left({\mu^2-E_{1,\kappa}^2}\right)^{\frac{1}{2}}}\right], ~~~~x=r^{\frac{1}{2}}$$$$
\footnotesize G_{1,\kappa}(r)\sim x^{2\kappa}(x-x_1)\exp\left[-x^2(\mu^2-E_{1,\kappa}^2)^{\frac{1}{2}}+\frac{2a_1 x(\mu+E_{1,\kappa})}{\left({\mu^2-E_{1,\kappa}^2}\right)^{\frac{1}{2}}}\right]\times~~~~~~~~~~~~~~~~ $$$$~~~~~~~~~~~~~~~~~~~~~~~~~\frac{1}{\mu+E_{1,\kappa}}\left\{-x^2(\mu^2-E_{1,\kappa}^2)^{\frac{1}{2}}+x\left[\frac{a_1 (\mu+E_{1,\kappa})}{\left({\mu^2-E_{1,\kappa}^2}\right)^{\frac{1}{2}}}+\frac{1/2}{x-x_1}\right]+2\kappa+1\right\}, 
\end{equation}
subject to the constraint
\begin{equation}\label{eq:25}
8(\mu+E_{1,\kappa})\left[a_3-\frac{\left(\kappa+\frac{3}{4}\right)a_1}{(\mu^2-E_{1,\kappa}^2)^{\frac{1}{2}}}\right]=-4\left(\mu^2-E_{1,\kappa}^2\right)^\frac{1}{2}x_1+4\kappa+3,
\end{equation}
where the roots $\{x_i\}$ satisfy the equation
\begin{equation}\label{eq:26}
x_1^2(\mu^2-E_{1,\kappa}^2)^{\frac{1}{2}}-\frac{a_1x_1(\mu+E_{1,\kappa})}{({\mu^2-E_{1,\kappa}^2})^{\frac{1}{2}}}-(\kappa+3/4)=0,
\end{equation}
It is important to note that a careful selection of $a_1$ and $a_2$ is crucial for the existence of bound state. Moreover, it is obvious from  Eq.\,\eqref{eq:23} that the energy can either be positive or negative depending on the values of parameters $a_1$ and $a_2$. For a given value of $\mu$, the energy moves from positive to negative states as  $a_1$ and $a_2$ increases while for any given $\mu$, $a_1$ and $a_2$, the energy moves from negative to positive states as the quantum number $n_r+2(\kappa+1)$ increases. Thus it is interesting to investigate the condition for zero-point energy state. To do this, we set $E_{n_r,\kappa}=0$ in  Eq.\,\eqref{eq:18} and obtain the critical condition
\begin{equation}\label{eq:26a}
(n'+2a_2)\mu=a_1^2,~~n'=n_r+2(\kappa+1).
\end{equation}
As example, we set $\mu=1$, $a_1=a_2=1$ and solve Eqs.\,\eqref{eq:18} and \eqref{eq:19}  simultaneously to explicitly obtain the energy $E_{n_r\kappa}$ and allowed values of the parameter $a_3$ for $n_r=0,1,2$ as shown in table 1. We note the degenerate states $|n_r,\kappa+1\rangle$ and $|n_r+2,\kappa\rangle$.

\begin{figure}[h]
\centering
\includegraphics[scale=.6]{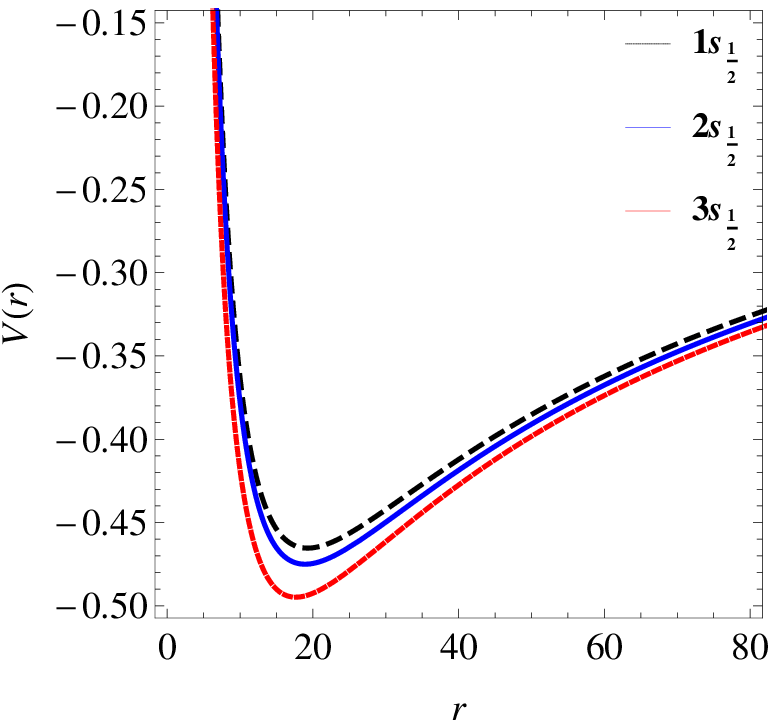}
\includegraphics[scale=.6]{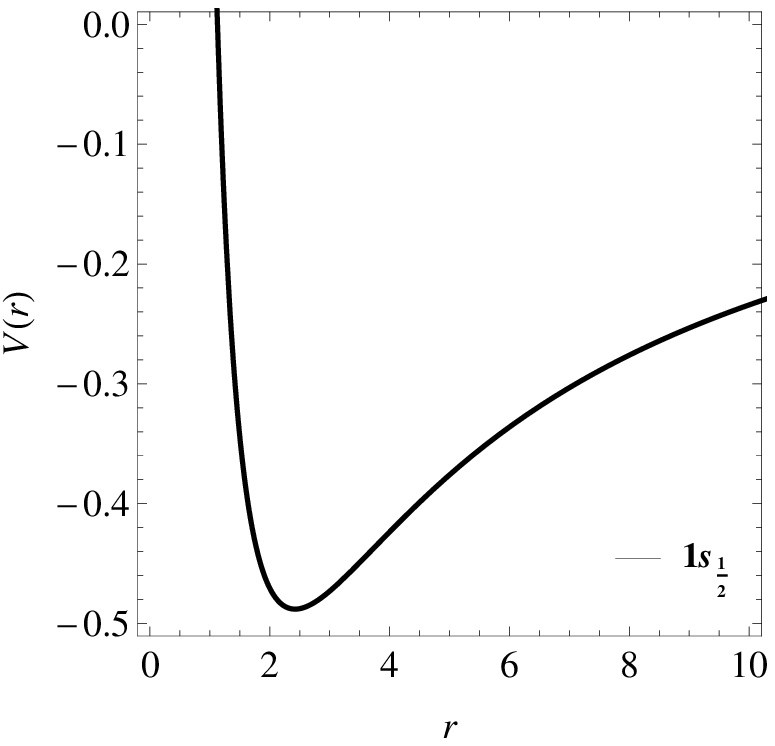}\\
\includegraphics[scale=.57]{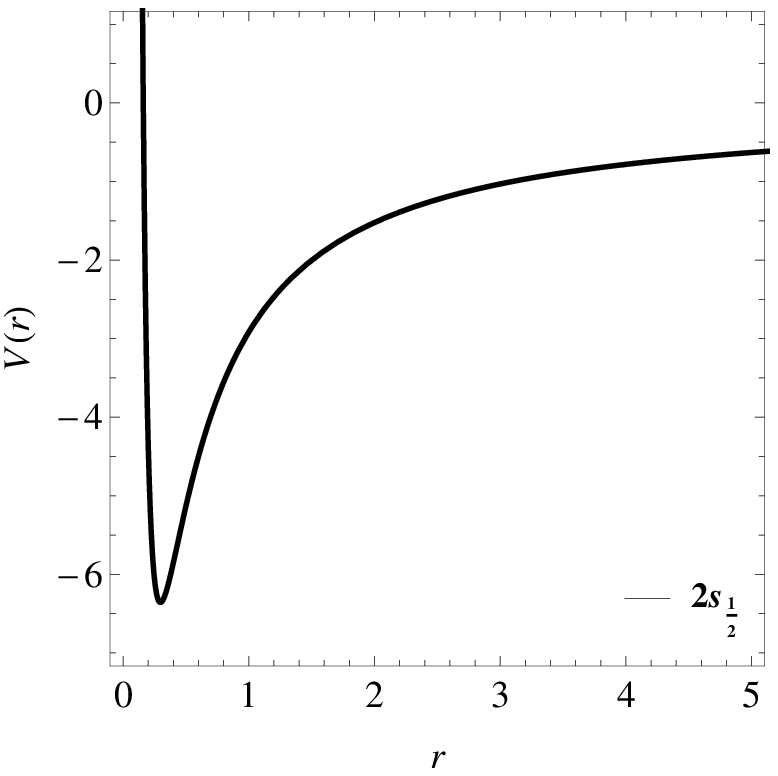}
\includegraphics[scale=.6]{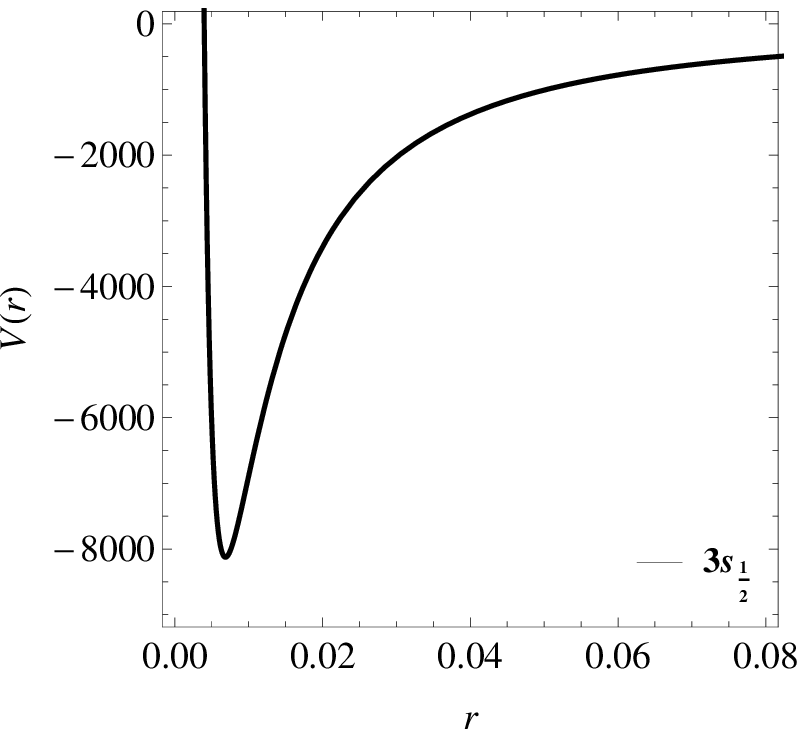}
\caption{Figure (a) shows the curves of the effective potential $V_{\mbox{\footnotesize eff}}(r)$ for $V_2(r)$ for states 1$s_{1/2}$, 2$s_{1/2}$ and 3$s_{1/2}$ with $a_1=a_2=1$ and $\mu=1$. Figures (b)-(d) show the curves of the effective potential $V_{\mbox{\footnotesize eff}}(r)$ for $V_3(r)$ for states 1$s_{1/2}$, 2$s_{1/2}$ and 3$s_{1/2}$ with $a_1=a_2=a_3=2$ and $\mu=1$}
\label{fig:energy}
\end{figure}

\section{Third-root singular potential ($N=3$)}
We now consider the potential of the form 
\begin{equation}\label{eq:r1}
V_{3}(r)=-\frac{a_1}{r^{\frac{1}{3}}}+\frac{a_2}{r^{\frac{2}{3}}}-\frac{a_3}{r}+\frac{a_4}{r^{\frac{4}{3}}}+\frac{a_5}{r^{\frac{5}{3}}},~~a_1,a_2,a_3>0
\end{equation}
thus taking the transformation 
\begin{equation}\label{eq:r2}
F_{n_r\kappa}(r)=x^{3(\kappa+1)}~e^{b_3x^3+b_2x^2+b_1x}u(x),~~~~x=r^{\frac{1}{3}},
\end{equation} and using  Eq.\,\eqref{eq:12}, we have
\begin{equation}\label{eq:r3}
xu''(x)+2\left[3b_3x^3+2b_2x^2+b_1x+3\kappa+2\right]u'(x)+\left\{9(E_{n_r\kappa}^2-\mu^2+b_3^2)x^5+[12b_2b_3+18(\mu+E_{n_r\kappa})a_1]x^4\right.$$$$
\left.+[4b_2^2+6b_1b_3-18(\mu+E_{n_r\kappa})a_2]x^3+[18b_3(\kappa+1)+4b_1b_2+18(\mu+E_{n_r\kappa})a_3]x^2\right.
$$$$\left.+\left[b_1^2+2b_2(6\kappa+5)-18(\mu+E_{n_r\kappa})a_4\right]x+2b_1(3\kappa+2)-18(\mu+E_{n_r\kappa})a_5\right\}u(x)=0
\end{equation}
 Eq.\,\eqref{eq:r3} is quasi-exactly solvable provided coefficients of the terms $x^5,x^4, x^3$ are zero. This yields the values of the parameters
\begin{equation}\label{eq:r4}
b_3=-\left(\mu^2-E_{n_r\kappa}^2\right)^{\frac{1}{2}},~~~b_2=\frac{3(\mu+E_{n_r\kappa})a_1}{2\left(\mu^2-E_{n_r\kappa}^2\right)^{\frac{1}{2}}}, ~~~b_1=-\frac{3(\mu+E_{n_r\kappa})a_2}{\left(\mu^2-E_{n_r\kappa}^2\right)^{\frac{1}{2}}}+\frac{3(\mu+E_{n_r\kappa})^2a_1^2}{2\left(\mu^2-E_{n_r\kappa}^2\right)^{\frac{3}{2}}},
\end{equation}
and reduces  Eq.\,\eqref{eq:r3} to
\begin{equation}\label{eq:r5}
xu''(x)+2\left[3b_3x^3+2b_2x^2+b_1x+3\kappa+2\right]u'(x)+\left\{[18b_3(\kappa+1)+4b_1b_2+18(\mu+E_{n_r\kappa})a_3]x^2\right.$$$$+[b_1^2+2b_2(6\kappa+5)-18(\mu+E_{n_r\kappa})a_4]x
\left.+2b_1(3\kappa+2)-18(\mu+E_{n_r\kappa})a_5\right\}u(x)=0.
\end{equation}

In this case, we seek the polynomial solution of the form
\begin{equation}\label{eq:r6}
u(x)=\prod_{m=1}^{n_r}(x-x_m),~~~~~~~~u(x)\equiv 1~~\mbox{for}~~n_r=0, 
\end{equation}
where $\{x_m\}$ are the roots of the polynomials to be determined, followed by the Bethe ansatz procedure used in section 2, we obtain the energy equation
\begin{equation}\label{eq:r7}
\left[n_r+3(\kappa+1)\right]\left(\mu^2-E_{n_r\kappa}^2\right)^\frac{1}{2}-\frac{3a_1^3(\mu+E_{n_r\kappa})^3}{2(\mu^2-E_{n_r\kappa}^2)^2}+\frac{3a_1a_2(\mu+E_{n_r\kappa})^2}{\mu^2-E_{n_r\kappa}^2}-3(\mu+E_{n_r\kappa})a_3=0,
\end{equation}
and the corresponding spinor wavefunction
\begin{equation}\label{eq:r8}\footnotesize
F_{n_r\kappa}(r)=x^{3(\kappa+1)}\left[\prod_{m=1}^{n_r}(x-x_m)\right]\hspace{4in}$$$$~~~~~~~~~~~~~~~~~~\times\exp\left\{-x^3(\mu^2-E_{n_r\kappa}^2)^{\frac{1}{2}}+\frac{3a_1 x^2(\mu+E_{n_r\kappa})}{2\left({\mu^2-E_{n_r\kappa}^2}\right)^{\frac{1}{2}}}-\left[\frac{3(\mu+E_{n_r\kappa})a_2}{\left(\mu^2-E_{n_r\kappa}^2\right)^{\frac{1}{2}}}-\frac{3(\mu+E_{n_r\kappa})^2a_1^2}{2\left(\mu^2-E_{n_r\kappa}^2\right)^{\frac{3}{2}}}\right]x
\right\}, ~~~~x=r^{\frac{1}{3}}$$$$
\footnotesize G_{n_r\kappa}(r)=x^{3\kappa}\left[\prod_{m=1}^{n_r}(x-x_m)\right]\hspace{4.3in}$$$$~~~~~~~~~~~~~~~~~~\times\exp\left\{-x^3(\mu^2-E_{n_r\kappa}^2)^{\frac{1}{2}}+\frac{3a_1 x^2(\mu+E_{n_r\kappa})}{2\left({\mu^2-E_{n_r\kappa}^2}\right)^{\frac{1}{2}}}-\left[\frac{3(\mu+E_{n_r\kappa})a_2}{\left(\mu^2-E_{n_r\kappa}^2\right)^{\frac{1}{2}}}-\frac{3(\mu+E_{n_r\kappa})^2a_1^2}{2\left(\mu^2-E_{n_r\kappa}^2\right)^{\frac{3}{2}}}\right]x
\right\}~~~~~~~~~~~~~~~~ $$$$~~~~~~~~~~~~~~~~~~~~~~~~~\times\frac{1}{\mu+E_{n_r\kappa}}\left(-x^3(\mu^2-E_{n_r\kappa}^2)^{\frac{1}{2}}+\frac{a_1 x^2(\mu+E_{n_r\kappa})}{\left({\mu^2-E_{n_r\kappa}^2}\right)^{\frac{1}{2}}}-\left[\frac{(\mu+E_{n_r\kappa})a_2}{\left(\mu^2-E_{n_r\kappa}^2\right)^{\frac{1}{2}}}-\frac{(\mu+E_{n_r\kappa})^2a_1^2}{2\left(\mu^2-E_{n_r\kappa}^2\right)^{\frac{3}{2}}}\right]x\right. $$$$\left.+\frac{1}{3}\sum_{m=1}^{n_r}\frac{x}{x-x_m}+2\kappa+1\right), 
\end{equation}
provided $a_4$ and $a_5$ take the values
\begin{equation}\label{eq:r9}
a_4=\frac{1}{\left(\mu+E_{n_r\kappa}\right)}\left[\frac{b_1^2}{18}+\frac{b_2}{9}\left(2n_r+6\kappa+5\right)+\frac{b_3}{3}\sum_{m=1}^{n_r}x_m\right],$$$$
a_5=\frac{1}{\left(\mu+E_{n_r\kappa}\right)}\left[\frac{b_1}{9}(n_r+3\kappa+2)+\frac{b_3}{3}\sum_{m=1}^{n_r}x_m^2-\frac{2b_2}{9}\sum_{m=1}^{n_r}x_m\right]
\end{equation}
respectively, where $b_1,b_2$ and $b_3$ are given in  Eq.\,\eqref{eq:r4}, with the roots $\{x_m\}$ satisfying the algebraic equations
\begin{equation}\label{eq:r9b}
\sum_{s\neq m}^{n_r}\frac{1}{x_m-x_s}+3b_3x_m^2+2b_2x_m+b_1+\frac{3\kappa+2}{x_m}=0,\hspace{0.2in}m=1,\dots, n_r.
\end{equation}
As examples of the general solution, the ground state solutions, which corresponds to $n_r=0$ is given by
\begin{equation}\label{eq:r10}
3(\kappa+1)\left(\mu^2-E_{0,\kappa}^2\right)^\frac{1}{2}-\frac{3a_1^3(\mu+E_{0,\kappa})^3}{2(\mu^2-E_{0,\kappa}^2)^2}+\frac{3a_1a_2(\mu+E_{0,\kappa})^2}{\mu^2-E_{0,\kappa}^2}-3(\mu+E_{0,\kappa})a_3=0,
\end{equation}
\begin{equation}\label{eq:r11}\footnotesize
F_{0,\kappa}(r)=x^{3(\kappa+1)}\exp\left\{-x^3(\mu^2-E_{0,\kappa}^2)^{\frac{1}{2}}+\frac{3a_1 x^2(\mu+E_{0,\kappa})}{2\left({\mu^2-E_{0,\kappa}^2}\right)^{\frac{1}{2}}}-\left[\frac{3(\mu+E_{0,\kappa})a_2}{\left(\mu^2-E_{0,\kappa}^2\right)^{\frac{1}{2}}}-\frac{3(\mu+E_{0,\kappa})^2a_1^2}{2\left(\mu^2-E_{0,\kappa}^2\right)^{\frac{3}{2}}}\right]x
\right\}, ~~~~x=r^{\frac{1}{3}}$$$$
\footnotesize G_{0,\kappa}(r)=x^{3\kappa}\exp\left[-x^3(\mu^2-E_{0,\kappa}^2)^{\frac{1}{2}}+\frac{3a_1 x^2(\mu+E_{0,\kappa})}{2\left({\mu^2-E_{0,\kappa}^2}\right)^{\frac{1}{2}}}-\left[\frac{3(\mu+E_{0,\kappa})a_2}{\left(\mu^2-E_{0,\kappa}^2\right)^{\frac{1}{2}}}-\frac{3(\mu+E_{0,\kappa})^2a_1^2}{2\left(\mu^2-E_{0,\kappa}^2\right)^{\frac{3}{2}}}\right]x
\right]~~~~~~~~~~~~~~~~ $$$$~~~~~~~~~~~~~~~~~~~~~~~~~\times\frac{1}{\mu+E_{0,\kappa}}\left(-x^3(\mu^2-E_{0,\kappa}^2)^{\frac{1}{2}}+\frac{a_1 x^2(\mu+E_{0,\kappa})}{\left({\mu^2-E_{0,\kappa}^2}\right)^{\frac{1}{2}}}-\left[\frac{(\mu+E_{0,\kappa})a_2}{\left(\mu^2-E_{0,\kappa}^2\right)^{\frac{1}{2}}}-\frac{(\mu+E_{0,\kappa})^2a_1^2}{2\left(\mu^2-E_{0,\kappa}^2\right)^{\frac{3}{2}}}\right]x\right. $$$$\left.+2\kappa+1\right), 
\end{equation}
provided $a_4$ and $a_5$ take the values
\begin{equation}\label{eq:r12}
a_4=\frac{1}{\left(\mu+E_{0,\kappa}\right)}\left[\frac{b_1^2}{18}+\frac{b_2}{9}\left(6\kappa+5\right)+\frac{b_3}{3}\right],$$$$
a_5=\frac{1}{\left(\mu+E_{0,\kappa}\right)}\left[\frac{b_1}{9}(3\kappa+2)\right]
\end{equation}
respectively, with $b_1$ and $b_2$ given by  Eq.\,\eqref{eq:r4}. In a similar fashion, the first excited state corresponds to $n_r=1$, with solutions
 \begin{equation}\label{eq:r13}
(3\kappa+4)\left(\mu^2-E_{1,\kappa}^2\right)^\frac{1}{2}-\frac{3a_1^3(\mu+E_{1,\kappa})^3}{2(\mu^2-E_{1,\kappa}^2)^2}+\frac{3a_1a_2(\mu+E_{1,\kappa})^2}{\mu^2-E_{1,\kappa}^2}-3(\mu+E_{1,\kappa})a_3=0,
\end{equation}
\begin{equation}\label{eq:r14}\footnotesize
F_{1,\kappa}(r)=x^{3(\kappa+1)}(x-x_1)\hspace{5.5in}$$$$\times\exp\left\{-x^3(\mu^2-E_{1,\kappa}^2)^{\frac{1}{2}}+\frac{3a_1 x^2(\mu+E_{1,\kappa})}{2\left({\mu^2-E_{1,\kappa}^2}\right)^{\frac{1}{2}}}-\left[\frac{3(\mu+E_{1,\kappa})a_2}{\left(\mu^2-E_{1,\kappa}^2\right)^{\frac{1}{2}}}-\frac{3(\mu+E_{1,\kappa})^2a_1^2}{2\left(\mu^2-E_{1,\kappa}^2\right)^{\frac{3}{2}}}\right]x
\right\}, ~~~~x=r^{\frac{1}{3}}$$$$
\footnotesize G_{1,\kappa}(r)=x^{3\kappa}(x-x_1)\exp\left\{-x^3(\mu^2-E_{1,\kappa}^2)^{\frac{1}{2}}+\frac{3a_1 x^2(\mu+E_{1,\kappa})}{2\left({\mu^2-E_{1,\kappa}^2}\right)^{\frac{1}{2}}}-\left[\frac{3(\mu+E_{1,\kappa})a_2}{\left(\mu^2-E_{1,\kappa}^2\right)^{\frac{1}{2}}}-\frac{3(\mu+E_{1,\kappa})^2a_1^2}{2\left(\mu^2-E_{1,\kappa}^2\right)^{\frac{3}{2}}}\right]x
\right\}~~~~~~~~~~~~~~~~ $$$$~~~~~~~~~~~~~~~~~~~~~~~~~\times\frac{1}{\mu+E_{1,\kappa}}\left(-x^3(\mu^2-E_{1,\kappa}^2)^{\frac{1}{2}}+\frac{a_1 x^2(\mu+E_{1,\kappa})}{\left({\mu^2-E_{1,\kappa}^2}\right)^{\frac{1}{2}}}-\left[\frac{(\mu+E_{1,\kappa})a_2}{\left(\mu^2-E_{1,\kappa}^2\right)^{\frac{1}{2}}}-\frac{(\mu+E_{1,\kappa})^2a_1^2}{2\left(\mu^2-E_{1,\kappa}^2\right)^{\frac{3}{2}}}\right]x\right. $$$$\left.+\frac{x/3}{x-x_1}+2\kappa+1\right), 
\end{equation}
provided $a_4$ and $a_5$ take the values
\begin{equation}\label{eq:r15}
a_4=\frac{1}{\left(\mu+E_{1,\kappa}\right)}\left[\frac{b_1^2}{18}+\frac{b_2}{9}\left(6\kappa+7\right)+\frac{b_3x_1}{3}\right],$$$$
a_5=\frac{1}{\left(\mu+E_{1,\kappa}\right)}\left[\frac{b_1}{3}(\kappa+1)+\frac{b_3x_1^2}{3}-\frac{2b_2x_1}{9}\right]
\end{equation}
with $x_1$ satisfying the equation
\begin{equation}\label{eq:r16}
3b_3x_1^2+2b_2x_1+b_1+\frac{3\kappa+2}{x_1}=0.
\end{equation}
The fact that $a_4$ and $a_5$ are obtained in terms of other parameters is as a result of the quasi-exact solvability of the system. While these restrictions are necessary for the quantization of the spectrum, we note that they do not affect the generality of the model. In other word, these constraints simply indicate that $a_4$ and $a_5$ must have some specific values (depending on $b_i$, $\kappa$ and roots $x_i$) for the  Eq.\,\eqref{eq:r3} to be exactly solvable. Moreover, the values of $a_1,a_2$ and $a_3$ are important for the bound state to exist.

Thus a careful selection of $a_1=a_2= a_3=2$ and $\mu=1$ yields the numerical values for the energies for the generalized third-root singular model, as shown in table 2, with  energy degeneracies within states $|n_r,\kappa+1\rangle$ and $|n_r+3,\kappa\rangle$. In a similar fashion, the critical condition for zero-energy state ($E_{n_r\kappa}=0$) can be easily obtain from  Eq.\,\eqref{eq:r7} as
\begin{equation}\label{eq:r17}
[(n^*-3a_3)\mu-3a_1a_2]\mu=3a_1^2,~~~n^*=n_r+3(\kappa+1).
\end{equation}
\section{Non-relativistic bound states and expectation values}
Interestingly, within the limits $\kappa\rightarrow\ell$, $\mu+E_{n_r,\kappa}\rightarrow 1/2$ and $\mu^2-E_{n_r,\kappa}^2\rightarrow -E_{n_r,\ell}'$,  Eq.\,\eqref{eq:10a} can be written as 
\begin{equation}\label{eq:m1}
H_N'\phi(r)=E_{n_r,\ell}'\phi(r)
\end{equation}
where $H'$ is the non-relativistic effective Hamiltonian given by ($2\mu=\hbar=1$)
\begin{equation}\label{eq:m2}
H'_N=-\frac{d^2}{dr^2}+\frac{\ell(\ell+1)}{r^2}+\sum_{p=1}^{2N-1}\frac{a_p}{r^{p/N}},
\end{equation} $\phi(r)$ and $E'_{n_r,\ell}$ are wavefunction and  energy respectively. In particular, for the $N=2$ case, the non relativistic energies are obtainable as 
\begin{equation}\label{eq:m4}
\left[n_r+2\ell+2)\right]\sqrt{-E_{n_r\ell}'}-\frac{a_1^2}{4\sqrt{-E_{n_r\ell}'}}+a_2=0~~\Rightarrow~~E_{n,\ell}'=-\frac{1}{4n^2}\left[a_1^2n+2a_2^2+2a_2\sqrt{a_2^2+a_1^2n}\right]
\end{equation}
where we have defined $n=n_r+2\ell+2$. We note that from  Eq.\,\eqref{eq:m2}, if $N=1$, we have the Coulomb interaction whose energy levels $E_{n}'=-a_2^2/(n_r+\ell+1)^2$ are obtainable from  Eq.\,\eqref{eq:m4} when $a_1=a_3=0$. Moreover, the corresponding wavefunction are given as
\begin{equation}\label{eq:m5}\footnotesize
\phi_{n_r\ell}(r)\sim x^{2(\ell+1)}\left[\prod_{i=1}^{n_r}(x-x_i)\right]\exp\left[-x^2\sqrt{-E_{n_r\ell}'}-\frac{a_1 x}{\sqrt{-E_{n_r\ell}'}}\right], ~~~~x=r^{\frac{1}{2}}$$$$
\end{equation}
subject to the constraint
\begin{equation}\label{eq:m6}
a_3-\frac{\left(\ell+\frac{3}{4}\right)a_1}{\sqrt{-E_{n_r\ell}'}}=-\sqrt{-E_{n_r\ell}'}\sum^{n_r}_{i=1}x_i+n_r\left(\ell+\frac{3}{4}\right),
\end{equation}
where the roots $\{x_i\}$ satisfy the Bethe ansatz equations
\begin{equation}\label{eq:m7}
\sum_{j\neq i}^{n_r}\frac{1}{x_i-x_j}=2x_i\sqrt{-E_{n_r\ell}'}-\frac{a_1}{\sqrt{-E_{n_r\ell}'}}-\frac{(4\ell+3)}{2x_i},\hspace{0.2in}i=1,\dots, n_r.
\end{equation}
We note that  Eq.\,\eqref{eq:m4} is in agreement with  Eq.\,(9) of \cite{BDK95}; however no closed-form expression for the wavefunction and constraint were obtained in the previous work.  

Furthermore, since the Hamiltonian $H'_N$ for the system is a function of some parameters $q$ (say) then  by the Hellmann-Feynmann theorem (HFT)\cite{DA10,DA09,GH37,RPF39} we have 
\begin{equation}\label{eq:m8}
\frac{\partial E'_{n_r,\ell}(q)}{\partial q}=\langle\phi(q)|\frac{\partial H'_N(q)}{\partial q}|\phi(q)\rangle. 
\end{equation}
The effective Hamiltonian for case $N=2$ is given by
 \begin{equation}\label{eq:m9}
H'_2=-\frac{d^2}{dr^2}+\frac{\ell(\ell+1)}{r^2}-\frac{a_1}{\sqrt{r}}+\frac{a_2}{r}+\frac{a_3}{r^{\frac{3}{2}}},
\end{equation}
thus with the choice of $q=a_1$, we have
\begin{equation}\label{eq:m10}
\langle\phi(a_1)|\frac{\partial H'_2(a_1)}{\partial a_1}|\phi(a_1)\rangle=-\langle r^{-\frac{1}{2}}\rangle
\end{equation} and
\begin{equation}\label{eq:m11}
\frac{\partial E'_{n_r,\ell}(a_1)}{\partial a_1}=-\frac{a_1}{2n}\left[1+\frac{a_2}{\sqrt{a_2^2+a_1^2n}}\right].
\end{equation} Thus be HFT,
\begin{equation}\label{eq:m12}
\langle r^{-\frac{1}{2}}\rangle=\frac{a_1}{2n}\left[1+\frac{a_2}{\sqrt{a_2^2+a_1^2n}}\right].
\end{equation}
Similarly for $q=a_2$, we obatain
 \begin{equation}\label{eq:m13}
\langle r^{-1}\rangle=-\frac{1}{2n^2}\left[2a_2+\frac{a_2^2}{\sqrt{a_2^2+a_1^2n}}+\sqrt{a_2^2+a_1^2n}\right],
\end{equation}
and for $q=\ell$,
\begin{equation}\label{eq:m14}
\langle r^{-2}\rangle=-\frac{1}{n(2\ell+1)}\left[{2}E'_{n_r,\ell}+\frac{a^2_1}{4n}\left(1+\frac{a_2}{\sqrt{a_2^2+a_1^2n}}\right)\right].
\end{equation}
Finally, we note that a similar procedure can be followed to obtain the expectation values for the generalized third-root singular potential.
\section{Concluding Remarks}

In this paper, we have presented exact (Bethe ansatz) solutions to the Dirac equation for a class of singular fractional power potentials. For the first two members of this class of potentials, we showed that the Dirac equation is reducible to a quasi-exactly solvable differential equation which has exact solutions, provided the parameters satisfy certain constraints. We obtained closed form expressions for the energies and eigenfunctions in terms of the roots of Bethe ansatz equations. Similar degeneracy between the states $|n_r,\kappa+1\rangle$ and $|n_r+N,\kappa\rangle$ was observed for any member potential $V_N(r)$.

By taking appropriate limits, we discussed some non-relativistic properties of the models, and using the Hellmann-Feynmann theorem, explicit expressions for some expectation values were obtained. We hope that the present findings will lead to new applications for this class of fractional singular potentials.

\section*{Acknowledgments }
We wish to thank the referees for their very useful comments. This work was supported by the Australian IPRS
and a University of Queensland Centennial Scholarship. YZZ is supported in part by the Australian Research Council through Discovery Project DP110103434.
DA is indebted to Father J and Agboola B for their support during the preparation of the manuscript.
\begin{table}[h]
\caption{Exact relativistic energies for the square-root singular potential ($N=2$), with $\mu=1$ and free parameters $a_1=a_2=1$.}

\centering
\begin{tabular}{p{.2in}l p{.8in}l c c c c}\hline
$n_r$&$\kappa$&\hfil($\ell$, $j$)&$\hfil a_3$&$E_{n_r,\kappa}$\\\hline\\
0&1&\hfil 1$s_{1/2}$&2.531648042&$-\frac{1}{10}\,\sqrt [3]{13+10\,\sqrt {15}}+{\frac {11}{10}}\,{\frac {1}{\sqrt [3]{13+10\,\sqrt {15}}}}+\frac{4}{5}
$\\\\
&2&\hfil 1$p_{3/2}$&4.247765159&$-\frac{1}{20}\,\sqrt [3]{118+30\,\sqrt {35}}+{\frac {13}{10}}\,{\frac {1}{\sqrt [3]{118+30\,\sqrt {35}}}}+{\frac {9}{10}}
$\\\\
&3&\hfil 1$d_{5/2}$&6.107053526&$-\frac{1}{34}\,\sqrt [3]{433+204\,\sqrt {7}}+{\frac {47}{34}}\,{\frac {1}{
\sqrt [3]{433+204\,\sqrt {7}}}}+{\frac {16}{17}}
$\\\\
&4&\hfil 1$f_{7/2}$&8.087930004&$-{\frac {1}{52}}\,\sqrt [3]{1126+390\,\sqrt {11}}+{\frac {37}{26}}\,{
\frac {1}{\sqrt [3]{1126+390\,\sqrt {11}}}}+{\frac {25}{26}}
$\\\\
&5&\hfil 1$g_{9/2}$&10.17558343&$-{\frac {1}{74}}\,\sqrt [3]{2413+222\,\sqrt {143}}+{\frac {107}{74}}\,
{\frac {1}{\sqrt [3]{2413+222\,\sqrt {143}}}}+{\frac {36}{37}}
$\\\\
1&1&\hfil 2$s_{1/2}$&2.262287002&$-\frac{1}{29}\,\sqrt [3]{383+290\,\sqrt {6}}+{\frac {71}{29}}\,{\frac {1}{
\sqrt [3]{383+290\,\sqrt {6}}}}+{\frac {25}{29}}
$\\\\
&2&\hfil 2$p_{3/2}$&4.201005545&$-{\frac {1}{53}}\,\sqrt [3]{1919+1484\,\sqrt {3}}+{\frac {143}{53}}\,{
\frac {1}{\sqrt [3]{1919+1484\,\sqrt {3}}}}+{\frac {49}{53}}
$\\\\
&3&\hfil 2$d_{5/2}$&6.285744349&$\frac{4}{5}$\\\\
&4&\hfil 2$f_{7/2}$&8.494170316&$-{\frac {1}{125}}\,\sqrt [3]{13439+2750\,\sqrt {30}}+{\frac {359}{125}
}\,{\frac {1}{\sqrt [3]{13439+2750\,\sqrt {30}}}}+{\frac {121}{125}}
$\\\\
&5&\hfil 2$g_{9/2}$&10.81102582&$-{\frac {1}{173}}\,\sqrt [3]{26879+4498\,\sqrt {42}}+{\frac {503}{173}
}\,{\frac {1}{\sqrt [3]{26879+4498\,\sqrt {42}}}}+{\frac {169}{173}}
$\\\\
2&1&\hfil 3$s_{1/2}$&2.903513638&$-\frac{1}{20}\,\sqrt [3]{118+30\,\sqrt {35}}+{\frac {13}{10}}\,{\frac {1}{
\sqrt [3]{118+30\,\sqrt {35}}}}+{\frac {9}{10}}
$\\\\
&2&\hfil 3$p_{3/2}$&5.185092156&$-\frac{1}{34}\,\sqrt [3]{433+204\,\sqrt {7}}+{\frac {47}{34}}\,{\frac {1}{
\sqrt [3]{433+204\,\sqrt {7}}}}+{\frac {16}{17}}
$\\\\
&3&\hfil 3$d_{5/2}$&7.592941604&$-{\frac {1}{52}}\,\sqrt [3]{1126+390\,\sqrt {11}}+{\frac {37}{26}}\,{
\frac {1}{\sqrt [3]{1126+390\,\sqrt {11}}}}+{\frac {25}{26}}
$\\\\
&4&\hfil 3$f_{7/2}$&10.11211386&$-{\frac {1}{74}}\,\sqrt [3]{2413+222\,\sqrt {143}}+{\frac {107}{74}}\,
{\frac {1}{\sqrt [3]{2413+222\,\sqrt {143}}}}+{\frac {36}{37}}
$\\\\
&5&\hfil 3$g_{9/2}$&12.73121312&$-{\frac {1}{100}}\,\sqrt [3]{4558+350\,\sqrt {195}}+{\frac {73}{50}}\,
{\frac {1}{\sqrt [3]{4558+350\,\sqrt {195}}}}+{\frac {49}{50}}
$\\\hline

\end{tabular}

\label{tab:}
\end{table}

\begin{table}[h]
\caption{Exact relativistic energies for the third-root singular potential ($N=3$), with $\mu=1$ and free parameters $a_1=a_2=a_3=2$.}

\centering
\begin{tabular}{p{.2in}l p{.8in}l c c c c c}\hline
$n_r$&$\kappa$&\hfil($\ell$, $j$)&$\hfil a_4$&$a_5$&$E_{n_r,\kappa}$\\\hline\\
0&1&\hfil 1$s_{1/2}$&1.958333333&-0.8333333333&0\\\\
&2&\hfil 1$p_{3/2}$&2.960561941&-1.153142721&0.1242218925\\\\
&3&\hfil 1$d_{5/2}$&4.000970707&-1.400062497&0.2012486748\\\\
&4&\hfil 1$f_{7/2}$&5.072119166&-1.583630614&0.2558964562\\\\
&5&\hfil 1$g_{9/2}$&6.169422769&-1.709933223&0.2976807024\\\\
1&1&\hfil 2$s_{1/2}$&1.818578964&-2.135565865&0.04969611507\\\\
&2&\hfil 2$p_{3/2}$&2.798193832&-2.734248685&0.1533707648\\\\
&3&\hfil 2$d_{5/2}$&3.823365431&-3.211896333&0.2213237925\\\\
&4&\hfil 2$d_{5/2}$&4.882868288&-3.597655799&0.2709662658\\\\
&5&\hfil 2$g_{9/2}$&5.970673184&-3.906627168&0.3096166374\\\\
2&1&\hfil 3$s_{1/2}$&1.645358882&-3.718197892&0.09022131430\\\\
&2&\hfil 3$p_{3/2}$&2.505019545&-5.086724151&0.1787860494\\\\
&3&\hfil 3$d_{5/2}$&3.452841027&-6.224631833&0.2394325294\\\\
&4&\hfil 3$f_{7/2}$&4.456655401&-7.189499169&0.2848409032\\\\
&5&\hfil 3$g_{9/2}$&5.501066171&-8.021053784&0.3207569813\\\hline

\end{tabular}

\label{tab:}
\end{table}

\begin{figure}[h]
\centering
\includegraphics[scale=0.6]{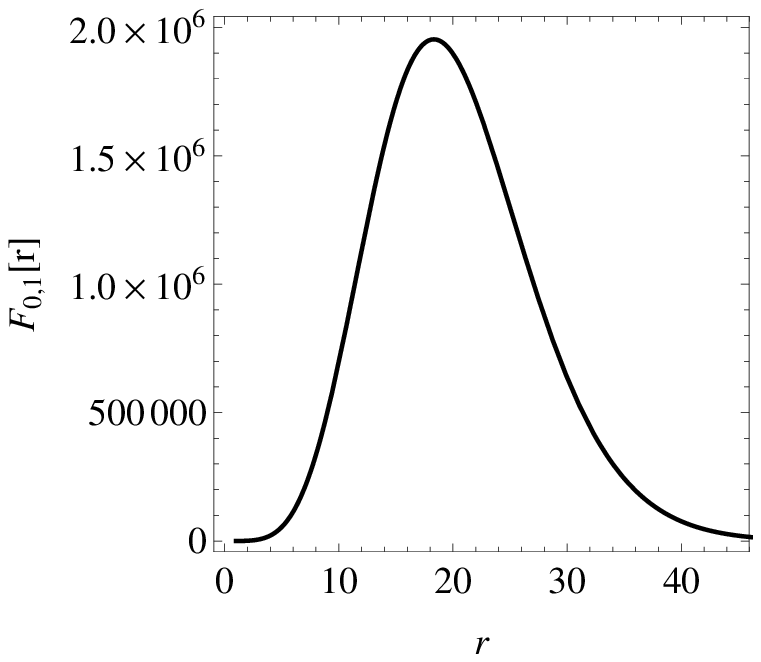}
\includegraphics[scale=.58]{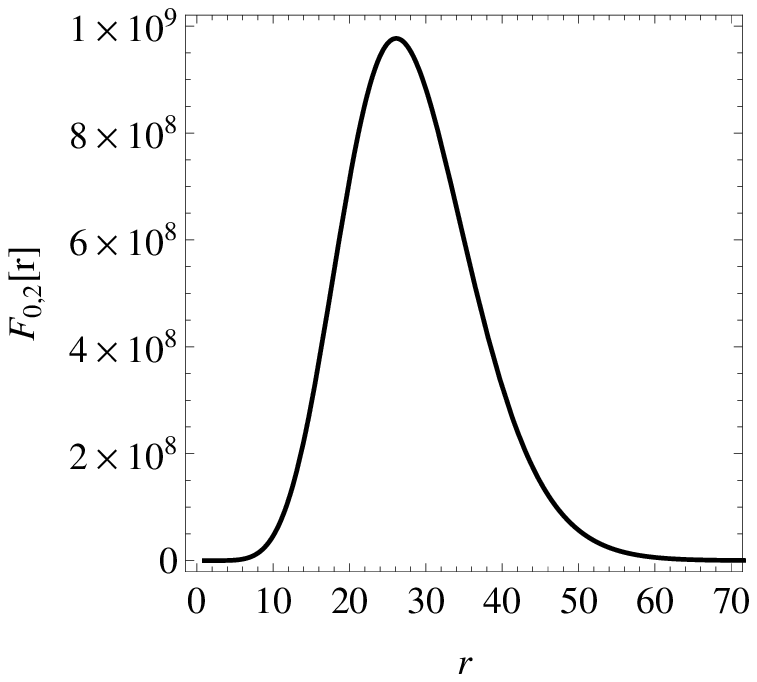}
\includegraphics[scale=.6]{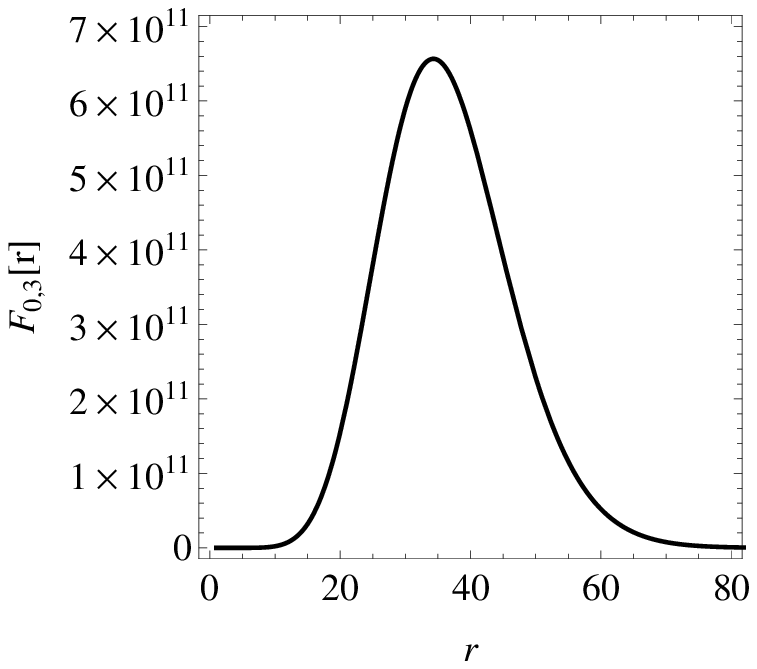}\\
\includegraphics[scale=.6]{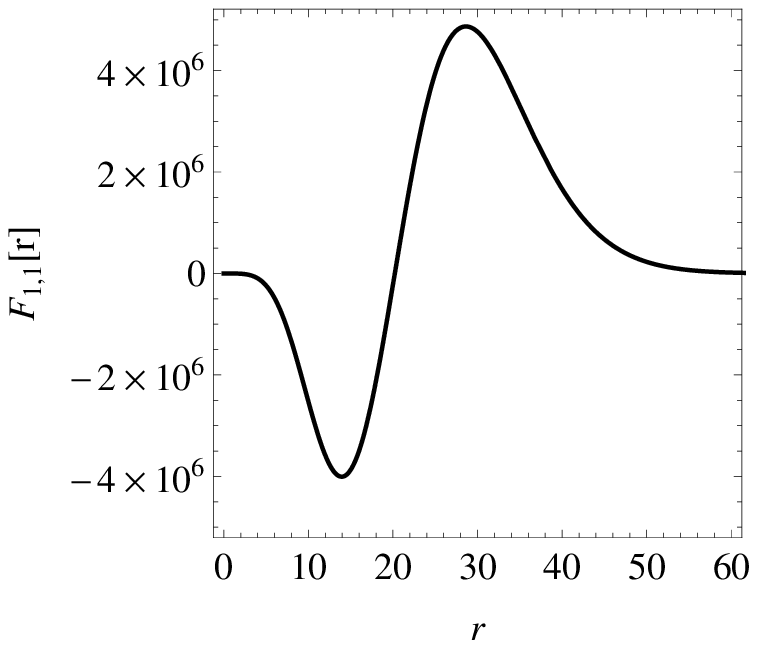}
\includegraphics[scale=.6]{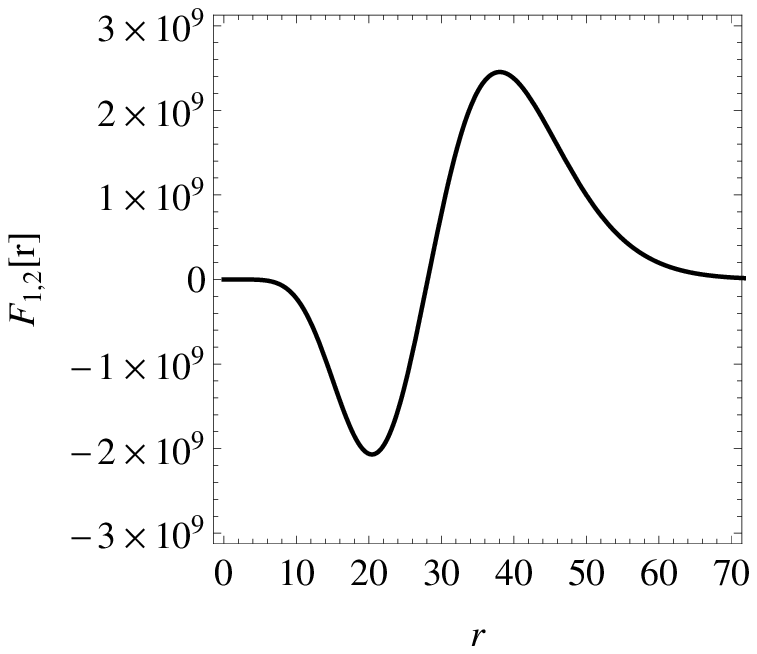}
\includegraphics[scale=.6]{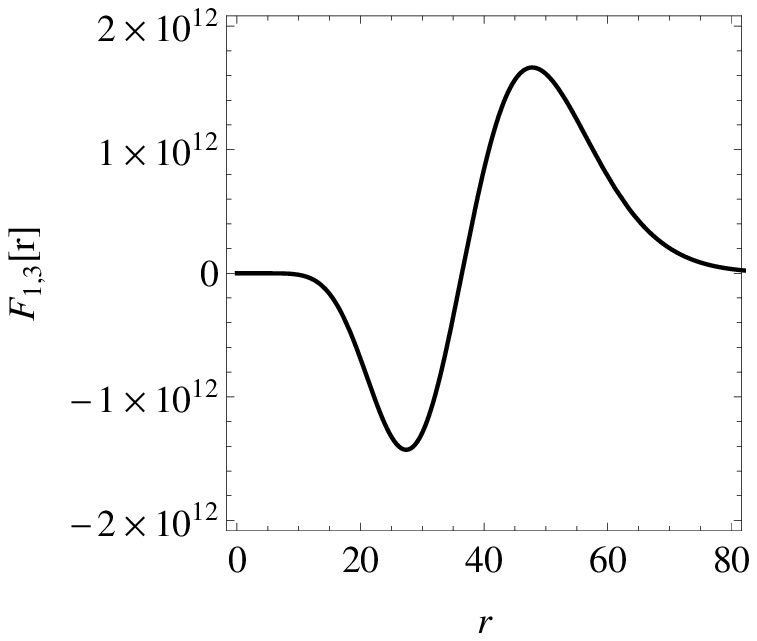}
\caption{The wavefunction $F_{n_r,\kappa}(r)$ for the ground and first excited states of the square-root singular potential ($N=2$) with parameters $a_1=a_2=1$ and $\mu=1$ and $\kappa=1-3$.}
\label{fig:energy}
\end{figure}

\begin{figure}[h]
\centering
\includegraphics[scale=.6]{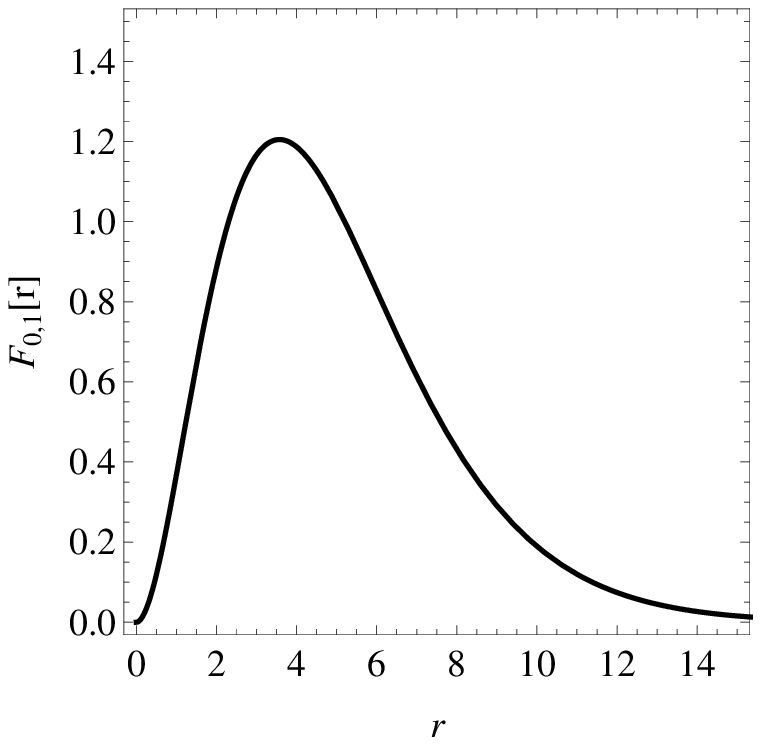}
\includegraphics[scale=.6]{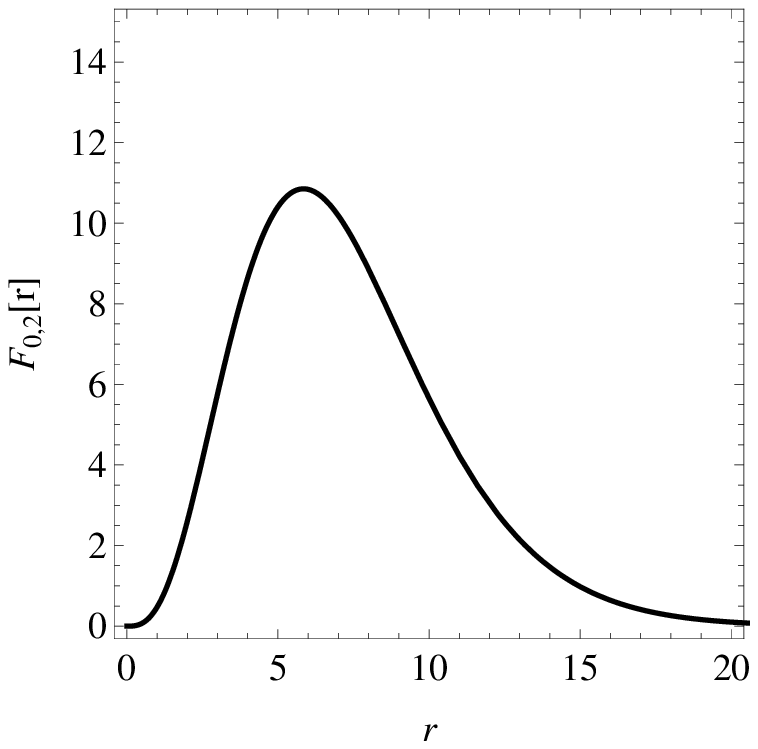}
\includegraphics[scale=.6]{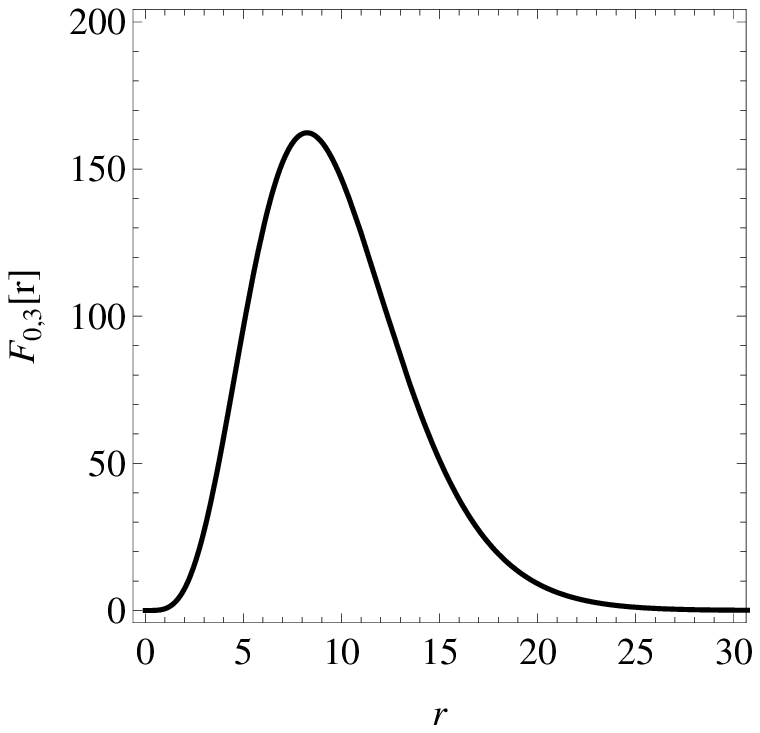}\\
\includegraphics[scale=.6]{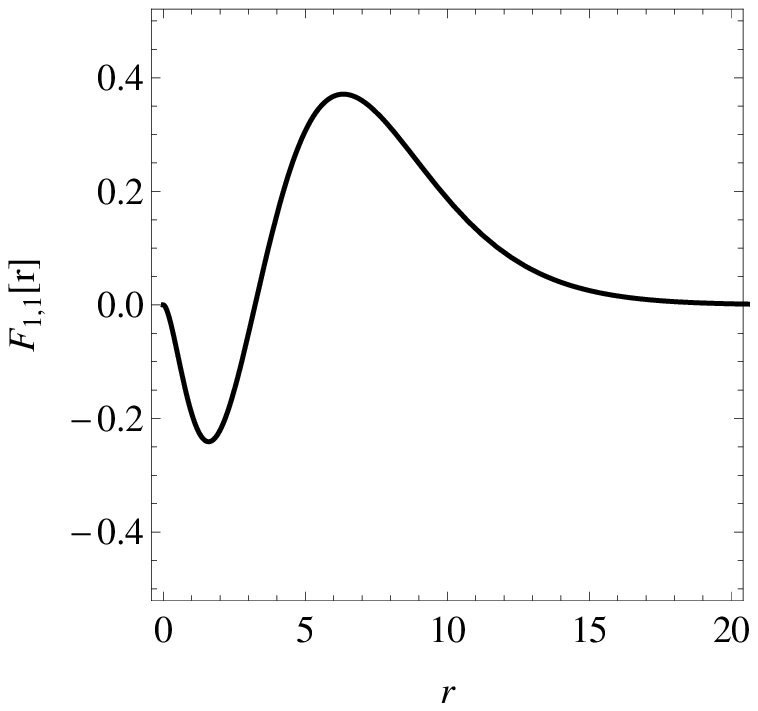}
\includegraphics[scale=.6]{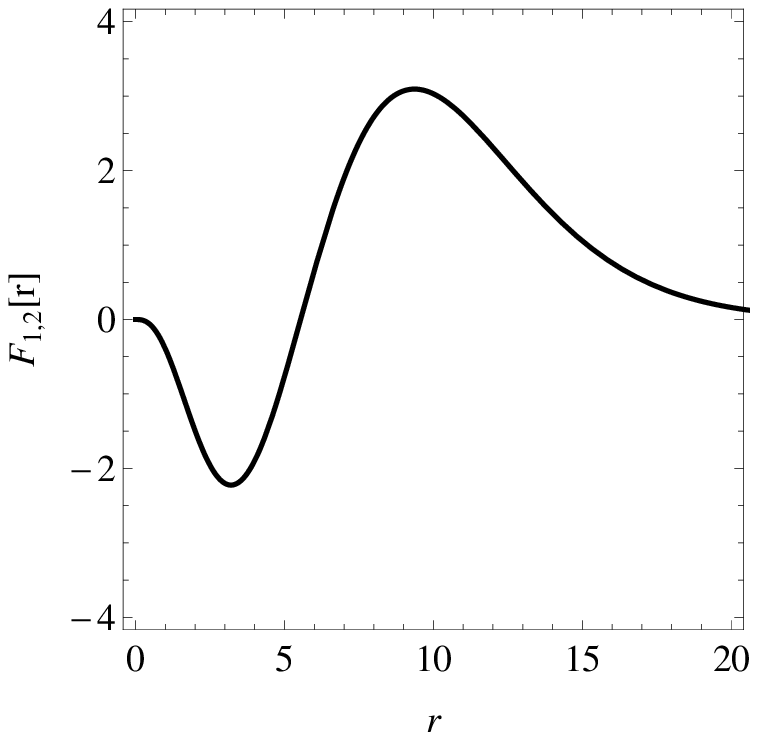}
\includegraphics[scale=.6]{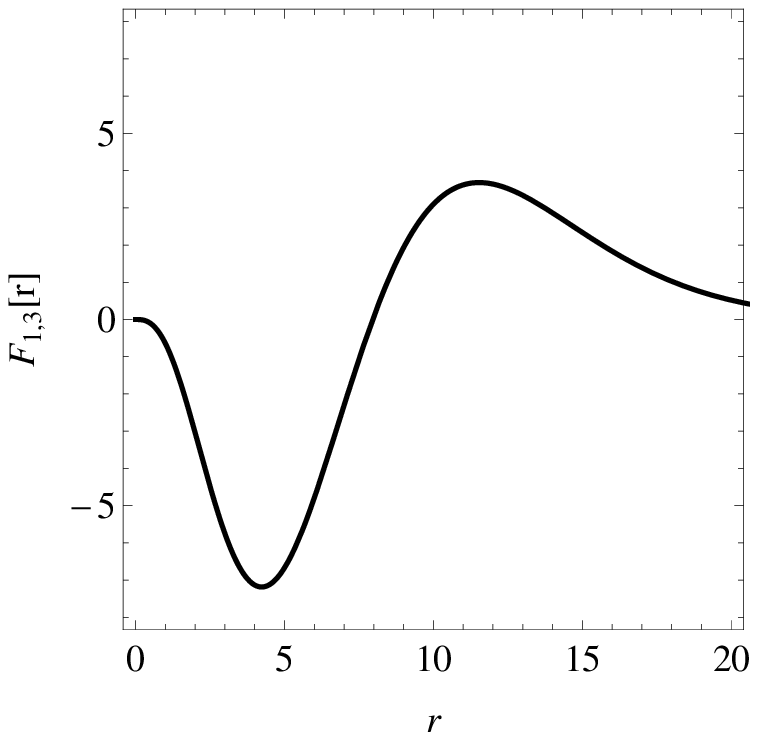}
\caption{The wavefunction $F_{n_r,\kappa}(r)$ for the ground and first excited states of the third-root singular potential ($N=3$) with parameters $a_1=a_2=a_3=1$ and $\mu=1$ and $\kappa=1-3$.}
\label{fig:energy}
\end{figure}

\end{document}